# Extraordinary efficient spin-orbit torque switching in (W, Ta)/epitaxial-Co$_{60}$Fe$_{40}$/TiN heterostructures

Nilamani Behera,[1] Rahul Gupta,[1] Sajid Husain[1], Vineet Barwal,[2]

Dinesh K. Pandya,[2] Sujeet Chaudhary,[2] Rimantas Brucas,[1] Peter Svedlindh,[1] and Ankit Kumar[1,*]

[1]*Department of Engineering Sciences, Uppsala University, Box 534, SE-751 21 Uppsala, Sweden*
[2]*Thin Film Laboratory, Department of Physics, Indian Institute of Technology Delhi,
New Delhi 110016, India*

The giant spin Hall effect in magnetic heterostructures along with low spin memory loss and high interfacial spin mixing conductance are prerequisites to realize energy efficient spin torque based logic devices. We report giant spin Hall angle (SHA) of 28.67 (5.09) for W (Ta) interfaced epi- Co$_{60}$Fe$_{40}$/TiN structures. The spin-orbit torque switching current density ($J_{Crit}$) is as low as 1.82 (8.21) MA/cm$^2$ in W (Ta)/Co$_{60}$Fe$_{40}$($t_{CoFe}$)/TiN structures whose origin lies in the epitaxial interfaces. These structures also exhibit very low spin memory loss and high spin mixing conductance. These extraordinary values of SHA and therefore ultra-low $J_{Crit}$ in semiconducting industry compatible epitaxial materials combinations open up a new direction for the realization of energy efficient spin logic devices by utilizing epitaxial interfaces.

Pure spin current based spintronic devices have advantages over conventional microelectronic devices owing to low energy dissipation, fast switching, and high-speed data processing etc., and can be integrated with microelectronic semiconductor devices for better functionality[1-4]. These new spintronic devices mainly work on the principle of spin manipulation employing the spin Hall effect (SHE) and the Rashba Edelstein effect (REE). The basic building block of the spin devices is comprised of ferromagnetic (FM)/ non-magnetic (NM) bilayers; and the NM layers and their interfaces should possess strong relativistic spin-orbit interaction (SOI). The SOI can be of bulk as well as of interfacial origin, and it generates spin-orbit torques (SOTs), i.e. damping-like (DL) and field-like (FL) SOTs[4-8]. A charge current applied to SHE and REE based devices generates a transverse spin current and therefore a spin orbit torque at the FM/NM interface, which can be used to manipulate the FM state[4-7]. In contrast, devices based on the inverse effects, the inverse SHE (ISHE) and inverse REE (IREE) convert a spin current generated by spin pumping into a charge current in the NM layer by the ISHE and at the FM/NM interface by the IREE[6-12]. Bulk SOI of the NM layer is responsible for the SHE and ISHE mechanisms[2, 4, 5, 8], while interfacial SOI is responsible for the REE and IREE mechanisms[6, 7, 9, 10]. Recently, a strong DL SOT in ferromagnetic semiconductor (Ga,Mn)As thin films[7] and Py/CuOx heterostructures[10] has been reported, the origin of which lies in crystal inversion asymmetry and interfacial SOI induced Berry curvature[7, 10], respectively. High values of the spin Hall angle, which is a measure of the SHE, have been reported, of the order of 100% in $Ni_{0.6}Cu_{0.4}$ [13], 1880 % in topological insulator $Bi_xSe_{1-x}$ [14], and 5200% in conducting topological insulator $Bi_{0.9}Sb_{0.1}$ [15] thin films. The topological insulators exhibit the highest value of spin Hall angle, however their thin film fabrication is challenging. Therefore, present technological focus is to find CMOS (complementary metal-oxide semiconductor) technology compatible material combinations with even larger SOTs, or more specifically larger spin angular momentum transfer across the interfaces in magnetic heterostructures. Spin manipulation in the bulk or at the interface plays a decisive role in building the next generation of spintronics devices, viz. spin torque magnetic random access memories (ST-MRAMs), spin logic devices, ST-transistors and ST nano-oscillators [4-10].

To optimize the bilayer stacks for devices, there is a need to control interfacial properties like spin backflow, spin memory loss and magnetic proximity effects that may arise at/near the interface in FM/NM bilayer structures, properties that demean the overall spin transport in the structures[3, 16-22]. It is known that the spin current at the interface also exhibits spin memory loss in the presence of interfacial SOI, by creating a parallel relaxation path for the spin current[3, 19-22]. The spin memory loss can be quantified as the relative amount of spin current which is dissipated while passing through the FM/NM interface. Since the heterostructures interfaces play a decisive role in controlling the spin angular momentum transfer, it would be intriguing to investigate in detail how interfacial parameters affect the spin Hall angle and therefore the spin orbit torque switching efficiency when NM layers are interfaced with epitaxial FM heterostructures.

In this work, spin pumping measurements were performed on β-W/epi-$Co_{60}Fe_{40}$/TiN and β-Ta/epi-$Co_{60}Fe_{40}$/TiN heterostructures to estimate precisely the interfacial parameters; spin mixing conductance ($g_{NM/CoFe}^{\uparrow\downarrow}$), spin memory loss (SML) and spin diffusion length ($\lambda_{SD}$). The interfacial parameters corrected spin Hall angle values of W and Ta interfaced epitaxial heterostructures have been determined by performing line-shape analysis and dc current induced modulation of the effective damping using data obtained from spin torque ferromagnetic resonance (STFMR) measurements. The dc bias dependent changes of the effective damping confirm the presence of significantly large interfacial DL torques in these heterostructures whose origin lie both at the W and Ta interfaces as well as at the TiN interface with epitaxial $Co_{60}Fe_{40}$. The critical switching current density of the W (Ta) interface at which the effective damping reverses sign is 1.82 (8.21) MA/cm$^2$, which is comparable to the values reported for conductive topological insulators[15].

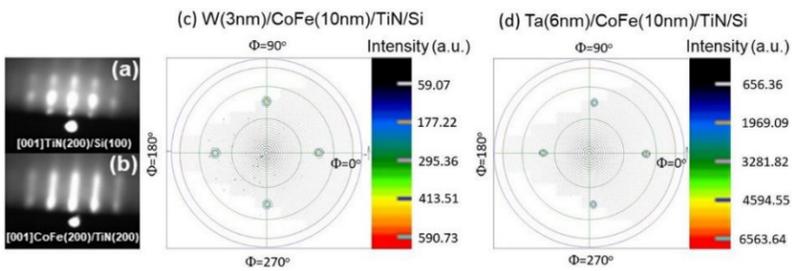

**Figure 1** The RHEED patterns along [001] of (a) TiN 10nm, and (b) $Co_{60}Fe_{40}$ 10nm thin films. (c) and (d) show the pole figure XRD patterns of $Co_{60}Fe_{40}$(022) for W/ epi-$Co_{60}Fe_{40}$/TiN/Si and Ta/epi-$Co_{60}Fe_{40}$/TiN/Si thin films, respectively, confirming the epitaxial quality of the $Co_{60}Fe_{40}$ thin films grown on TiN buffered Si (100) substrates.

**Structural Characterization**

Figure 1 (a) and (b) show the RHEED patterns recorded along the [001] direction for the TiN(200) film grown on a $2 \times 1$ reconstructed Si(100) substrate and for the $Co_{60}Fe_{40}$ film grown on TiN buffered Si(100), respectively. The evolution of elongated and sharp streaks confirms the 2D-epitaxial quality of the CoFe layers, giving evidence of a CoFe(200)[001]/TiN(200)[001]/Si(400)[001] orientation relationship of these samples. To further confirm the epitaxial growth of CoFe on the TiN buffered Si substrate, texture analyses were performed by measuring XRD pole figures. Figure 1 (c) and (d) show the pole figure XRD patterns of the CoFe(022) plane at 2θ=45.2° for the W/epi-CoFe/TiN/Si and Ta/epi-CoFe/TiN/Si thin films confirming the epitaxial quality of the CoFe thin films. The X-ray reflectivity (XRR) determined thickness of each individual layer is in close agreement with the nominal values predicted from the growth rates; the XRR results are presented in supplementary information (SI). The interface roughness of each individual layer is in the range of ≤ 1 nm, and these values are also closely matching with previously reported values [23-25].

**Spin Hall angle measurement**

In-plane angle dependent FMR measurements reveal that the epitaxially grown CoFe thin films exhibit cubic anisotropy, as shown in SI. In the *in-plane* FMR configuration, the resonance field $H_r$ and full-width at half-maximum linewidth $\Delta H$ of the spectra were determined by using a lock-in-amplifier based FMR technique; *in-plane* FMR spectra are presented in SI and the estimated parameters are shown in Table 1. The *out-of-plane* FMR measurements were performed using vector network analyzer to determine two magnon scattering free $H_r$, and $\Delta H$; see SI. The *out-of-plane* FMR estimated values of the spin mixing conductance, intrinsic Gilbert damping ($\alpha_0$), spin pumping ($\alpha_S$), spin memory loss and spin diffusion length in the W and Ta interfaced FM heterostructures are also shown in Table 1.

To determine the spin Hall angle and the spin-orbit torque switching efficiency, STFMR measurements were performed on patterned (20×100 μm$^2$) W(6nm)/epi-CoFe(10nm)/TiN/Si and Ta(6nm)/epi-CoFe(10nm)/TiN/Si thin films. Schematic figures of the STFMR setup and the torques acting on the magnetization are shown in Figs. 2 (a) and (b) (see Ref. 26 for measurement details). In the STFMR measurements, the microwave current $I_{rf}$ generates an Oersted field and a transverse spin current in the NM layer due to the SHE. The torques due to the Oersted field ($\tau_{Oe}$) and the transverse spin current ($\tau_{AD}$) contribute with anti-symmetric and symmetric profiles, respectively, to the FMR line-shape. The anti-damping $\tau_{AD}$ torque acts against the intrinsic damping torque $\tau_D$ in the FM layer. In case of interfacial SOI, the Rashba Edelstein effect contributes with a field-like torque ($\tau_{FL}$) with direction opposite to that of $\tau_{Oe}$. At resonance, the temporal variation of the magnetization vector with respect to the direction of $I_{rf}$ produces due to the anisotropic magnetoresistance of the FM layer a time varying resistance that mixes with $I_{rf}$ yielding a dc voltage output. In our case, using a low-frequency (1 kHz) modulation of $I_{rf}$ the STFMR signal is detected using a lock-in amplifier. The observed STFMR spectra exhibit a combination of symmetric and anti-symmetric Lorentzian weight factors[27-33]. The STFMR spectrum can be expressed as

$$V_{mix} = V_0 \left[ S F_S(H) + A F_A(H) \right], \quad (1)$$

|  |  | $\mu_0 M_s$(T) | $\alpha_0$(×10⁻³) | $\alpha_s$(×10⁻²) (nm) | $g_{NM/CoFe}^{\uparrow\downarrow}$ (×10¹⁹) m⁻² | $\lambda_{SD}$(nm) |
|---|---|---|---|---|---|---|
| W/epi-Co₆₀Fe₄₀/TiN/Si | in-plane FMR | 2.46±0.03 | 4.95±0.23 | 2.48±0.15 |  |  |
|  | out-of-plane FMR | 2.34±0.03 | 3.50±0.09 | 2.45±0.08 | 3.60±0.02 | 3.20±0.90 |
| Ta/epi-Co₆₀Fe₄₀/TiN/Si | in-plane FMR | 2.40±0.02 | 3.54±0.12 | 1.80±0.11 |  |  |
|  | out-of-plane FMR | 2.27±0.01 | 3.34±0.04 | 0.74±0.04 | 1.13±0.02 | 6.50±0.75 |
|  |  |  |  | $\mu_0 M_{eff}$(T) | $|\theta_{SH}^{LS}|$ | $|\theta_{SH}^{MOD}|$ |
| W(6nm)/epi-Co₆₀Fe₄₀(10nm)/TiN/Si |  | STFMR |  | 2.08±0.01 | 33.68 | 28.67 |
| Ta(6nm)/epi-Co₆₀Fe₄₀(10nm)/TiN/Si |  | STFMR |  | 2.14±0.01 | 4.57 | 5.09 |

**Table 1** Parameters determined from *in-plane*, *out-of-plane* FMR and STFMR measurements on W/epi-Co₆₀Fe₄₀/TiN/Si and Ta/epi-Co₆₀Fe₄₀/TiN/Si thin films. The observed difference in the intrinsic damping $\alpha_0$ values measured by *in-plane* and *out-of-plane* FMR is due to the presence of two-magnon scattering contribution in the *in-plane* FMR determined values.

where $V_0$ is the amplitude of the mixing voltage. $S = \hbar J_S/2e\mu_0 M_s t_{CoFe}$ and $A = (H_{rf} + H_{FL})\sqrt{(1 + \frac{M_{eff}}{H_r})}$ are symmetric and anti-symmetric Lorentzian weight factors, accounting for anti-damping and field-like torques, respectively. $\hbar J_S/2e$ is the spin current density generated in the NM layer in units of J/m², $H_{FL}$ represents the field-like torque, and $\mu_0 M_{eff}$ is the effective magnetization. $F_S(H) = \left(\frac{\Delta H}{2}\right)^2/\left[\left(\frac{\Delta H}{2}\right)^2 + (H - H_r)^2\right]$ is the symmetric and $F_A(H) = F_S(H) \times \frac{(H-H_r)}{\left(\frac{\Delta H}{2}\right)}$ is the anti-symmetric Lorentzian function (for details see Ref. 28). The Oersted field generated by $I_{rf}$ can be expressed as $H_{rf} = \frac{t_{NM}}{2}J_{C,rf}$, where $J_{C,rf}$ is the microwave current density in the NM layer. The linear change of $S$ with increasing microwave power for $\leq 13$ dBm power shows that the experimental results are not influenced by heating at the power used in the measurements (13 dBm); see SI for details.

The spin Hall angle ($\theta_{SH}^{LS}$) can be calculated from the expression[27-33]

$$\theta_{SH}^{LS} = \frac{J_S}{J_{C,rf}} = \frac{S}{A}\frac{2e\mu_0 M_s t_{CoFe}}{\hbar}\left(\frac{t_{NM}}{2} + \frac{H_{FL}}{J_{C,rf}}\right)\sqrt{\left(1 + \frac{M_{eff}}{H_r}\right)},$$

(2)

The line-shape parameters $S$, $A$, $H_r$ and $\Delta H$ were obtained by fitting the STFMR spectra using Eq. (1); results of the fitting are shown in Fig. 2 (c-d). The results of $f$ vs. $H_r$ and $\Delta H$ vs. $f$ are shown in Fig. 2 (e) and (f), respectively, together with fits employing the *in-plane* Kittel equation, $f = \frac{\mu_0\gamma}{2\pi}[(H_r + H_K)(H_r + H_K + M_{eff})]^{\frac{1}{2}}$ [34] and $\mu_0\Delta H = \mu_0\Delta H_0 + $ $(4\pi\alpha_{eff}f/\gamma)$ [35], where $\gamma$=180 GHz/T is the gyromagnetic ratio, and $\alpha_{eff}$ is the effective Gilbert damping of the heterostructure. The fitted values of $\mu_0 M_{eff}$ are 2.08±0.01T and 2.14±0.01T for W(6nm)/epi-Co₆₀Fe₄₀(10nm)/TiN/Si and Ta(6nm)/epi-Co₆₀Fe₄₀(10nm)/TiN/Si, respectively, which closely match with the FMR determined values. The determined values of $\alpha_{eff}$ are found to be 0.0073±0.0002 and 0.0056±0.0001 for W(6nm)/epi-CoFe(10nm)/TiN/Si and Ta(6nm)/epi-CoFe(10nm)/TiN/Si, respectively. Since the intrinsic damping $\alpha_0$ values of CoFe in both samples are nearly equal (cf. Fig. S6(c) of SI), the higher value of $\alpha_{eff}$ in W(6nm)/epi-CoFe(10nm)/TiN/Si is due to a larger spin pumping contribution. The inhomogeneity contribution to damping $\Delta H_0$ is found to be 5mT and 1mT for W(6nm)/epi-CoFe(10nm)/TiN/Si and Ta(6nm)/epi-CoFe(10nm)/TiN/Si, respectively. To estimate the true $\theta_{SH}^{LS}$ values of the W and Ta interfaced heterostructures, Eq. (2) must be corrected for the spin pumping contribution in the symmetric line-shape, and the TiN layer Oersted and interface field-like contributions in the anti-symmetric line-shape.

During spin transfer torque or spin pumping measurements, the precessing magnetization in the FM layer transfers a pure spin current to the NM layer in the FM/NM system. Using Eq. (2) to determine $\theta_{SH}^{LS}$, only accounts for the spin-orbit torque, Oersted and interface field-like contributions from the W/Ta layer in the line-shape. The SOT contribution weight factor in the STFMR spectrum can be expressed as $\eta_{SOT} = 1/(1 + \frac{V_{ISHE}}{V_{STFMR}^{sym}})$, where $V_{ISHE}$ and $V_{STFMR}^{sym}$ are the spin-pumping and SOT contributions, respectively, to the total symmetric part of the resonance line-shape ($V_{Total}^{sym} = V_{ISHE} + V_{STFMR}^{sym}$).

The frequency dependent $\eta_{SOT}$ values are calculated by using the method presented in Ref. 26 and *out-of-plane* FMR determined parameters values (See SI for details), and subsequently the interfacial parameters corrected spin pumping

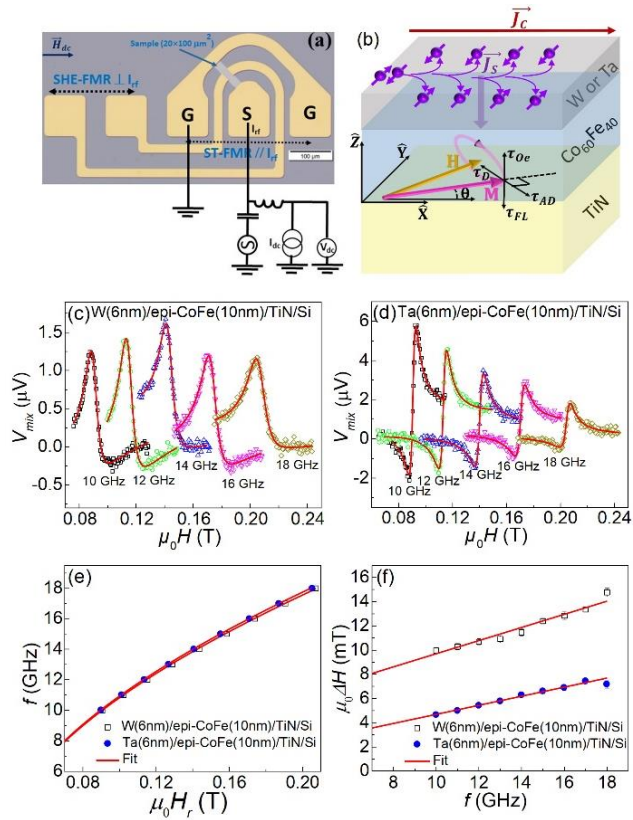

**Figure 2** (a) Schematic of the patterned structure and the STFMR setup. (b) Schematic of torque induced precession of the magnetization $M$ around its equilibrium direction, where θ is the angle of $M$ with respect to the XY plane. $\tau_D$, $\tau_{AD}$, $\tau_{Oe}$ and $\tau_{FL}$ are intrinsic damping torque, spin current anti-damping-like torque, Oersted field-like torque, and Rashba Edelstein field-like torque, respectively. (c) and (d) STFMR spectra in the frequency range 10-18 GHz of W(6nm)/epi-Co$_{60}$Fe$_{40}$(10nm)/TiN/Si and Ta(6nm)/epi-Co$_{60}$Fe$_{40}$(10nm)/TiN/Si thin films, respectively. The red solid lines are the respective fits to the experimental data. (e) $f$ vs. $\mu_0 H_r$ and (f) $\mu_0 \Delta H$ vs. $f$ obtained from STFMR measurements for W(6nm)/epi-Co$_{60}$Fe$_{40}$(10nm)/TiN/Si and Ta(6nm)/epi-Co$_{60}$Fe$_{40}$(10nm)/TiN/Si thin films at $I_{dc} = 0$ mA. The red solid lines are fits to the experimental data (see main text for explanations).

contributions ($V_{ISHE}$) are subtracted from the STFMR spectra. The W and Ta layer Oersted field contribution weight factor in the STFMR spectrum can be expressed as $\Lambda_{NM} = 1/(1 + \frac{A_{TiN}}{A_{NM}})$, where $A_{TiN}$ and $A_{NM}$ are the TiN and W or Ta layers antisymmetric contributions, respectively, to the total anti-symmetric part of the resonance line-shape ($A_{Total}^{Asym} = A_{TiN} + A_{NM}$); details are presented in SI. The spin-pumping and TiN Oersted and interface field-like corrected average values (using Eq. (S11) in SI) of $|\theta_{SH}^{LS}|$ are 33.68 and 4.57 for the W/epi-CoFe(10nm)/TiN/Si and Ta/epi-CoFe(10nm)/TiN/Si heterostructures, respectively, which clearly reflects the impact of accounting for the TiN layer when estimating the spin Hall angle. STFMR measurements have also been performed on a epi-CoFe(10nm)/TiN/Si heterostructure to estimate the CoFe(10nm)/TiN interface spin Hall angle. However, the TiN STFMR spectrum peak-to-peak amplitude is small compared to the amplitude of the Ta and W based heterostructure recorded spectra; see SI for details. We have estimated the spin Hall angle by correcting the line-shape of the STFMR spectrum for extrinsic contributions. However, we emphasize that these corrected values of the spin Hall angle only provide approximate information relating to the spin Hall effect and spin-orbit torque switching efficiency of the W and Ta interfaced epi-CoFe(10nm)/TiN/Si structures; hence, need to confirm using a method which is free from extrinsic contributions.

In order to make a more reliable determination of the spin Hall angle and spin-orbit torque switching efficiency, one can use the dc induced change of the damping, often referred to as the modulation of damping (MOD) method, for which the spin pumping (ISHE) and field-like contributions are absent. In the MOD method, one measures the change of the STFMR linewidth and the effective damping parameter as a charge current $I_{dc}$ is applied to the patterned structure[27-29]. The $I_{dc}$ induced change of the effective Gilbert damping $\alpha_{eff}(I_{dc})$ is given by[28,29]

$$\alpha_{eff}(I_{dc}) - \alpha_{eff}(I_{dc} = 0) = \left(\frac{\sin\varphi}{(H_r + 0.5 M_{eff})\mu_0 M_s t_{CoFe}} \frac{\hbar}{2e}\right) J_s,$$

(3)

where the spin current density $J_S = \theta_{SH}^{MOD} J_{C,dc}(NM)$;

$$J_{C,dc}((W,Ta) + TiN) = I_{dc}\left[\left(\frac{1}{A_{W,Ta}} \frac{R_{CoFe} \times R_{TiN}}{(R_{W,Ta} \times R_{TiN}) + (R_{CoFe} \times R_{TiN}) + (R_{W,Ta} \times R_{CoFe})}\right) + \left(\frac{1}{A_{TiN}} \frac{R_{CoFe} \times R_{W,Ta}}{(R_{W,Ta} \times R_{TiN}) + (R_{CoFe} \times R_{TiN}) + (R_{W,Ta} \times R_{CoFe})}\right)\right]$$

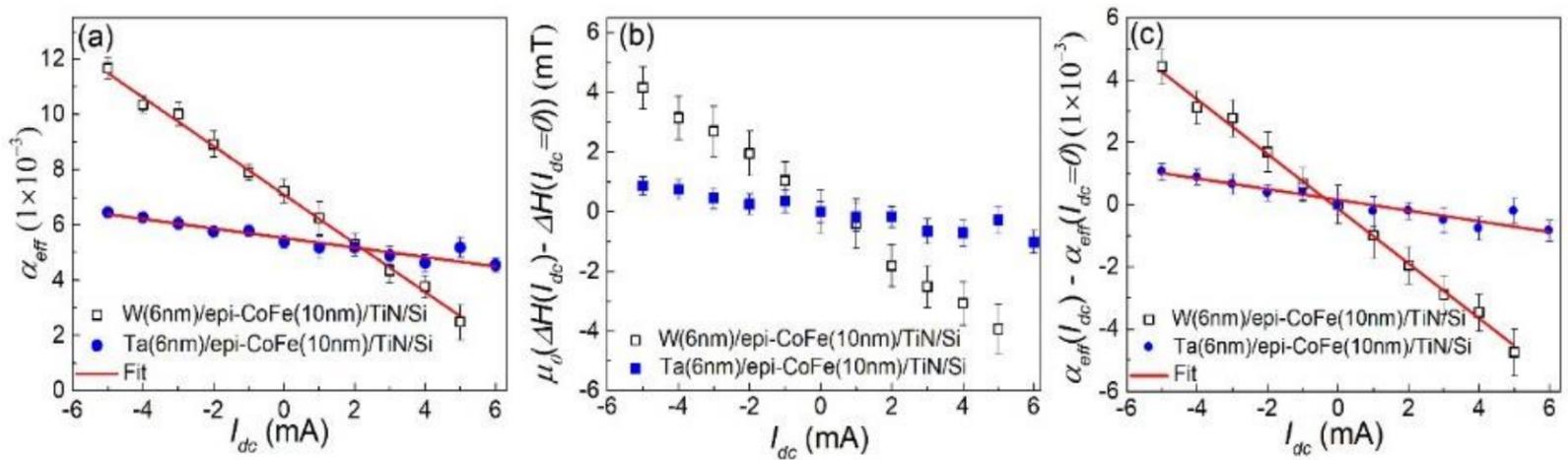

**Figure 3** (a) Effective damping parameter $\alpha_{eff}$ vs. $I_{dc}$ for positive applied fields. (b) $\Delta H(I_{dc}) - \Delta H(I_{dc} = 0)$ at $f$ =13 GHz, and (c) $\alpha_{eff}(I_{dc}) - \alpha_{eff}(I_{dc} = 0)$ vs. $I_{dc}$ for positive applied fields. The solid lines are linear fits to the experimental data.

Here $J_{C,dc}$ is the charge current density in the NM layers. $A_{NM}$ is the cross sectional area of the NM layer and $R_{NM}$ is the resistance of the NM layer. The measured resistivity values of W, Ta, TiN and CoFe are 283 μΩ-cm, 183 μΩ-cm, 98 μΩ-cm, and 72 μΩ-cm, respectively. The charge current in the W (Ta) layer is about 7% (10%) of $I_{dc}$. These values are close to the value reported by Huang et al. for the charge current in Ta (6.8% of $I_{dc}$) in Ta(5nm)/CoFeB(1.5-4 nm)/Pt(5nm) heterostructures[36]. However, Nan et al.[37] reported a charge current corresponding to 34.5% of $I_{dc}$ in the Ta layer in Ta(6nm)/Py(4nm) structures. Since our structures include a TiN seed layer with resistivity 98 μΩ-cm, a significant part of $I_{dc}$ is expected to go through the TiN layer. The spin Hall angle $\theta_{SH}^{MOD}$ can be estimated by measuring the $I_{dc}$ dependent rate of change of the effective damping, $\partial \alpha_{eff}(I_{dc})/\partial I_{dc}$. Hence $\theta_{SH}^{MOD}$ can be expressed as[29]

$$\theta_{SH}^{MOD} = \left[ \frac{\partial \alpha_{eff}/\partial J_{C,dc}(NM)}{\left(\frac{\sin\varphi}{(H_r + M_{eff})\mu_0 M_s t_{CoFe}} \frac{\hbar}{2e}\right)} \right], \quad (4)$$

STFMR spectra were recorded for different $I_{dc}$ in the range +5 mA to −5mA on W(6nm)/epi-CoFe(10nm)/TiN/Si and Ta(6nm)/epi-CoFe(6nm)/TiN/Si patterned structures. The recorded spectra were fitted using Eq. (1) to determine the line-shape parameters, and subsequently $\alpha_{eff}(I_{dc})$ was extracted for both the structures; the results are shown in Fig. 3(a). $\alpha_{eff}(I_{dc})$ increases from 0.0025±0.0006 to 0.0117±0.0004 for W(6nm)/epi-CoFe(10nm)/TiN/Si as $I_{dc}$ is varied from +5mA to −5mA, and from 0.0047±0.0025 to 0.0067±0.0019 for Ta(6nm)/epi-CoFe(10nm)/TiN/Si as $I_{dc}$ is varied from +6mA to 5mA. To make this data more understandable, the change in $\Delta H$ and $\alpha_{eff}$ with $I_{dc}$, i.e. $\Delta H(I_{dc}) - \Delta H(I_{dc} = 0)$ and $\alpha_{eff}(I_{dc}) - \alpha_{eff}(I_{dc} = 0)$ vs. $I_{dc}$, are plotted in Fig. 3(b) and (c). A linear decrease in $\Delta H(I_{dc}) - \Delta H(I_{dc} = 0)$ with increasing $I_{dc}$ for both samples clearly depicts the absence of heating in our measurements. The observed variation in $\alpha_{eff}(I_{dc})$ is larger in W(6nm)/epi-CoFe(10nm)/TiN/Si, which is due to the larger spin Hall angle and therefore higher SOT in the W based heterostructure. The percentage current-induced change of $\alpha_{eff}(I_{dc})$, defined as $\frac{\alpha_{eff}(I_{dc}=0) - \alpha_{eff}(I_{dc})}{\alpha_{eff}(I_{dc}=0)} \times 100\%$, at $I_{dc} = \pm 5mA$ is 65% and 22% for the W and Ta based heterostructures, respectively. Here, $I_{dc} = 5mA$ corresponds to a dc current density of $J_{dc} = 2.91 \times 10^9 \frac{A}{m^2}$ and $4.35 \times 10^9 \frac{A}{m^2}$ in the W and Ta layer, respectively. The observed current-induced change of $\alpha_{eff}(I_{dc})$ in W/epi-CoFe/TiN/Si is significantly higher than that reported by Pai et al.[31], where a current-induced change of 0.17% of the effective damping constant at $J_{dc} = \pm 1.0 \times 10^9 \frac{A}{m^2}$ was reported for the W(6nm)/$Co_{40}Fe_{40}B_{20}$(5nm)/$SiO_x$/Si system. Our current-induced change of $\alpha_{eff}(I_{dc})$ is significantly higher than that reported by Liu et al.[27]; they achieved a current-induced change of $\alpha_{eff}(I_{dc})$ of 0.003% at $J_{dc} = \pm 8.95 \times 10^9 \frac{A}{m^2}$ in Py(4nm)/Pt(6nm) structures. The calculated value of $|\theta_{SH}^{MOD}|$ averaged over all measured frequencies for the W(Ta)/epi-CoFe/TiN heterostructure using Eq. (4) is 28.67 (5.09). Here we notice that the role of the TiN interface cannot be ignored. These values are analogous to the values determined by line-shape analysis. However, ignoring the part of the current density passing through the TiN layer in the spin Hall angle determination, the values for W and Ta are 110.84 and 14.61, respectively.

The observed large $I_{dc}$ induced change of the effective damping parameter in W/epi-CoFe/TiN/Si is due to the SOI splitting of doubly degenerated tungsten bands near the Fermi level, resulting in a large spin Berry curvature and hence a large intrinsic SHE in the W based heterostructure[38]. In principle, the calculation of the spin Hall angle either by line-shape analysis, using Eq. (2), or the $I_{dc}$ dependent change of the effective damping parameter using Eq. (3), should give the same result. If no extra interfacial torques are acting on the CoFe layer, then $\theta_{SH}^{LS} = \theta_{SH}^{MOD}$. The observed small inconsistency in the experimentally determined spin Hall angle values of respective structure either by line-shape or by current induced change of the effective damping clearly indicates the presence of additional interfacial torques in the W(Ta)/CoFe(10nm)/TiN/Si structure. The critical switching current density of the W(Ta)/epi-CoFe/TiN heterostructure at which effective damping reverses sign is 1.82 (8.21) MA/cm$^2$, while it is 0.47 (2.87) MA/cm$^2$ for the W(Ta) interface alone.

The determined ultra-low value of the critical current density is comparable to the value reported in conducting topological structures; Bi$_{0.9}$Sb$_{0.1}$/MnGa bilayers[15]. Recently, Gao et al.[10] have reported an intrinsic Berry curvature anti-damping-like torque in Py/CuO$_x$ heterostructures arising solely from interfacial SOI in the structure. Moreover, external field free switching was demonstrated in CoFeB,NiFe(4nm)/Ti(3nm)/CoFeB(1-1.4nm)/MgO(1.6nm) employing interfacial torques[39]. In these structures, the presence of non-negligible interface-generated spin currents stem from the FM(CoFeB,NiFe)/Ti and Ti/CoFeB interfaces[39]. Since TiN generates a non-negligible spin current at the CoFe/TiN interface, we find that the FM/TiN interface plays a significant role in the $I_{dc}$ dependent change of the effective damping, originating from anti-damping-like torques due to interfacial symmetry breaking and SOI. The interfacial spin torques are very sensitive to the crystallographic structure and therefore the orbital ordering at the interface, hence the epitaxial nature of the heterostructures appears to be responsible for the large switching efficiency. The observed extraordinary switching efficiency in the studied epitaxial heterostructures leaves an open question about the physics at epitaxial interfaces. Therefore, to conclude with certainty on the origin of the enhanced damping change with $I_{dc}$ will require detailed theoretical analysis; however, further experimental interfacial spin band structure investigations are key to unveil the potential of the studied structures for ultralow powered spin-logic circuits.

**Conclusions**

In summary, spin pumping and spin transfer torque studies have been performed on β-W/epi-CoFe(10nm)/TiN/Si and β-Ta/epi-CoFe(10nm)/TiN/Si heterostructures. This study clearly indicates the importance of correcting for extrinsic parameters to accurately determine interfacial parameters as well as the parameters describing the NM layer. The giant spin Hall angle and therefore the extraordinary spin-orbit torque switching efficiency in the studied semiconducting technology compatible W(Ta) and TiN interfaced FM heterostructures open up a new avenue for the realization of ultralow powered spintronic devices by utilizing epitaxial magnetic heterostructures.

**Materials and Methods**

Four sets of W and Ta interfaced epitaxial Co$_{60}$Fe$_{40}$ films were deposited on TiN(200)[100](10nm)/Si substrates by pulsed dc magnetron sputtering technique. In the first two sets of films, layers of W($t_W$) and Ta($t_{Ta}$) with different thickness in the range of 1-15 nm were deposited on epi-CoFe(10nm)/TiN/Si. In last two sets of films, two series of W(4nm)/epi-Co$_{60}$Fe$_{40}$ ($t_{CoFe}$)/TiN/Si, and Ta(5nm)/epi-Co$_{60}$Fe$_{40}$ ($t_{CoFe}$)/TiN/Si structures were grown; here $t_{CoFe}$ was varied in the range 3-17nm. It was also noted that the grown W and Ta layers mostly exhibit the desired β-phase (i.e. A-15 cubic for W and tetragonal for Ta) as reported in our previous work[40-41]. The details of the experimental methods are presented in SI file.

*Magn. Magn. Mater.* 444, 256–262(2017).

**Acknowledgements**

This work is supported by the Swedish Research Council (VR), grant no 2017-03799, and Olle Engkvist Byggmästare, project number 182-0365. Prof. Mikael Ottosson is acknowledged for help during XRD measurements.

**Authors Contributions**

*Correspondence and request for materials should be addressed to Email: <u>ankit.kumar@angstrom.uu.se</u>

**Competing financial interest**

The authors declares no competing financial interests.